\documentclass[prl,twocolumn,showpacs,superscriptaddress]{revtex4} 
\usepackage{graphicx}
\usepackage[dvips]{color}
\usepackage{colordvi}

\newcommand{\be}{\begin{eqnarray}}
\newcommand{\ee}{\end{eqnarray}}
\newcommand{\bes}{\begin{eqnarray*}}
\newcommand{\ees}{\end{eqnarray*}}

\begin{document} 
\title{Correlated Strength in Nuclear Spectral Function}  
\date{\today} 
\author{}\affiliation{}
\author{D.~Rohe}\email[]{Daniela.Rohe@unibas.ch}\affiliation{University of Basel,  CH--4056 Basel, Switzerland} 
\author{C.S.~Armstrong}\affiliation{Thomas Jefferson National Accelerator Facility, Newport News, VA 23606, USA} 
\author{R.~Asaturyan}\affiliation{Yerevan Physics Institute, Yerevan, Armenia}
\author{O.K.~Baker}\affiliation{Hampton University, Hampton, VA 23668, USA}
\author{S.~Bueltmann}\affiliation{University of Virginia, Charlottesville, VA 22903, USA}  
\author{C.~Carasco}\affiliation{University of  Basel,   CH--4056 Basel, Switzerland}
\author{D.~Day}\affiliation{University of Virginia, Charlottesville, VA 22903, USA}  
\author{R.~Ent}\affiliation{Thomas Jefferson National Accelerator Facility, Newport News, VA 23606, USA} 
\author{H.C.~Fenker}\affiliation{Thomas Jefferson National Accelerator Facility, Newport News, VA 23606, USA}
\author{K.~Garrow}\affiliation{Thomas Jefferson National Accelerator Facility, Newport News, VA 23606, USA} 
\author{A.~Gasparian}\affiliation{Hampton University, Hampton, VA 23668, USA}
\author{P.~Gueye}\affiliation{Hampton University, Hampton, VA 23668, USA}
\author{M.~Hauger}\affiliation{University of Basel,   CH--4056 Basel, Switzerland} 
\author{A.~Honegger}\affiliation{University of Basel,   CH--4056 Basel, Switzerland}
\author{J.~Jourdan}\affiliation{University of  Basel,   CH--4056 Basel, Switzerland} 
\author{C.E.~Keppel}\affiliation{Hampton University, Hampton, VA 23668, USA}
\author{G.~Kubon}\affiliation{University of  Basel,   CH--4056 Basel, Switzerland}
\author{R.~Lindgren}\affiliation{University of Virginia, Charlottesville, VA 22903, USA} 
\author{A.~Lung}\affiliation{Thomas Jefferson National Accelerator Facility, Newport News, VA 23606, USA}  
\author{D.J.~Mack}\affiliation{Thomas Jefferson National Accelerator Facility, Newport News, VA 23606, USA} 
\author{J.H.~Mitchell}\affiliation{Thomas Jefferson National Accelerator Facility, Newport News, VA 23606, USA} 
\author{H.~Mkrtchyan}\affiliation{Yerevan Physics Institute, Yerevan, Armenia} 
\author{D.~Mocelj}\affiliation{University of  Basel,   CH--4056 Basel, Switzerland}
\author{K.~Normand}\affiliation{University of  Basel,   CH--4056 Basel, Switzerland}
\author{T.~Petitjean}\affiliation{University of Basel,   CH--4056 Basel, Switzerland} 
\author{O.~Rondon}\affiliation{University of Virginia, Charlottesville, VA 22903, USA} 
\author{E.~Segbefia}\affiliation{Hampton University, Hampton, VA 23668, USA}
\author{I.~Sick}\affiliation{University of Basel,   CH--4056 Basel, Switzerland} 
\author{S.~Stepanyan}\affiliation{Yerevan Physics Institute, Yerevan, Armenia}  
\author{L.~Tang}\affiliation{Hampton University, Hampton, VA 23668, USA}
\author{F.~Tiefenbacher}\affiliation{University of Basel,   CH--4056 Basel, Switzerland} 
\author{W.F.~Vulcan}\affiliation{Thomas Jefferson National Accelerator Facility, Newport News, VA 23606, USA}
\author{G.~Warren}\affiliation{University of Basel,   CH--4056 Basel, Switzerland} 
\author{S.A.~Wood}\affiliation{Thomas Jefferson National Accelerator Facility, Newport News, VA 23606, USA}  
\author{L.~Yuan}\affiliation{Hampton University, Hampton, VA 23668, USA}
\author{M.~Zeier}\affiliation{University of Virginia, Charlottesville, VA 22903, USA}
\author{H.~Zhu}\affiliation{University of Virginia, Charlottesville, VA 22903, USA}
\author{B.~Zihlmann}\affiliation{University of Virginia, Charlottesville, VA 22903, USA}  

\collaboration{E97-006 Collaboration}\noaffiliation

\begin{abstract}
We have carried out an (e,e'p) experiment at high momentum transfer and in
parallel kinematics to measure the strength of the nuclear spectral function 
$S(k,E)$ at
high nucleon momenta $k$ and large removal energies $E$. This strength is related
to the presence of short-range and tensor correlations, and was known hitherto
only indirectly and with considerable uncertainty from the {\em lack} of 
strength in the independent-particle region. This experiment locates by direct
measurement the correlated strength predicted by theory.
\end{abstract} 
\pacs{21.10.Jx, 25.30.-c, 25.30.Fj}
\maketitle 

{\bf Introduction.} 
The concept of independent particle (IP) motion has been rather successful in the
description of atomic nuclei; the shell model, based on the assumption that 
nucleons  move,
independently from each other, in the average potential created by the
interaction with all other nucleons,  has been able to explain many 
nuclear properties. This success often comes at the expense of the need to use
effective operators that implicitly account for the shortcomings of the IP
basis.

A more fundamental approach to the understanding of nuclei has to start from the
underlying nucleon-nucleon (N-N) interaction. This N-N interaction is well known
from many experiments on N-N scattering, and several modern parameterizations
are available. The N-N interaction exhibits a strongly repulsive central
interaction at small internucleon distances, and at medium distances a strong 
tensor component. These features lead to properties of nuclear wave functions that are
beyond what is describable in terms of an IP model. In particular, strong 
short-range correlations (SRC) are expected to occur.

The effects of the short-range correlations were studied for systems where 
the Schr\"odinger equation can be solved for a realistic N-N interaction
\cite{Pandharipande97}. Very light nuclei (today up to A$\leq$10) and infinite nuclear
matter are amongst the systems where this is feasible 
\cite{Pieper01,Benhar89,Muther00}. The corresponding
calculations show that in a microscopic description of nuclear systems 
the short-range and tensor parts of
the N-N interaction have a very important, not to say dominating, influence
without which not even nuclear binding can be explained. 

The consequences of these short-range correlations are that the momentum 
distributions of 
nucleons acquire a tail extending to very high momenta $k$ and at the same 
time part of the strength, located in IP descriptions at low excitation energy 
$E$, is moved to very high excitation energies.

In the past, most experimental investigations were confined to rather low
momenta and energies, {\em i.e.} to the region where the strength is dominated
(but not entirely given) by the IP properties. In this region, the consequences
of short-range correlations are indicated primarily by a {\em depopulation} of states in
comparison to the predictions of IP models (including the long-range correlations
which can be described by configuration mixing). According to several 
calculations for infinite matter and finite nuclei using different realistic 
N--N potentials, a depopulation of the order of 15 to 20\% is expected.

Studies of  one-nucleon transfer reactions like (d,$^3$He) (see {\em e.g.} \cite{Kramer01})
 first appeared to be compatible with a 100\%
occupation of IP states (within $\pm$10MeV of the Fermi edge). It was later
shown by $(e,e'p)$ experiments that this result was a consequence of a biased
choice of the nucleon radial wave functions $R(r)$ used in the interpretation 
of the   surface-dominated transfer reactions \cite{Kramer01}. 
Reactions like $(e,e'p)$, 
which are more sensitive to the nuclear interior and  measure the nucleon
momentum distribution  as well,  
gave occupancies of IP states in the 60\%-70\% range, largely independent of
nuclear mass number A \cite{Lapikas93}. 
These ``occupancies'' 
correspond to  partial sums of spectroscopic factors up to a limit in
excitation energy. This limit was experimental and varied: it was set by the
demand that one must be able to still observe  and identify the
IP strength. At the same time, these and other \cite{Cavedon82,Donnelly84} 
experiments showed  that, apart from the overall normalization, 
the wave functions (in $k$ or $r$-space) have shapes quite similar  to the ones predicted
by IP models. 

From the experimental information available up to now, the  
depopulation of IP strength at low $k,E$ is unambiguous.  Determining  the
total  correlated strength is not so direct, however. The  total correlated
strength is a factor of 4 or so (see below)  smaller than the IP strength, 
and the determination of this strength by taking  the {\em difference} of the
experimental IP strength with unity suffers from the unfavorable propagation of
uncertainties in the experimental measurement and theoretical interpretation of the
$(e,e'p)$ data. A {\em direct} measurement of the correlated strength is needed.

{\bf Correlated strength from (e,e'p). }
According to calculations that solve the Schr\"odinger equation for a realistic
N-N interaction, the correlated
strength is expected to be identifiable at high nucleon momenta $k$ and high removal
energies $E$; there, the values of the nuclear spectral function $S(k,E)$, the
probability to find in the nucleus nucleons of given $k$ and $E$, is increased by
orders of magnitude relative to IP descriptions. The correlated strength also
contributes to the region dominated by the IP strength, but there it cannot be
isolated via (e,e'p). While initial searches for 
high-$k$ components \cite{Bobeldijk95b,Blomqvist95a} were restricted to 
low--lying states, it has been understood  for some time that the SRC   produce 
strength at high $k$ and $E$ {\em simultaneously} \cite{Dieperink76,Benhar89}.

Locating this strength at large $k$ {\em and} $E$  is difficult. 
The correlated strength (perhaps 20\%) is spread over a very large range in $E$
(one to several hundred $MeV$), so the density of $S(k,E)$ is very low.
Processes other than the single-step proton knockout  ---
the basis of the Plane Wave Impulse Approximation (PWIA) interpretation of 
$(e,e'p)$ --- can contribute.
Strength can be moved to large $E$ (appearing as large ``missing
energy'' $E_m$) by processes such as multi-nucleon knockout or
$\pi$-production, where the additional particle is not observed.
 Unless,  by the choice of
kinematics, this contribution can be reduced to a size where it can 
be corrected for by a calculation,
identification of the correlated strength is not possible.

A systematic study \cite{Sick97a} of $(e,e'p)$ data 
\cite{Marchand88,Weinstein90,Lourie86,Ulmer87,Baghaei89,Morrison93,Offerman96}
has shown that the best chance for an identification of the correlated strength
occurs for  data taken in {\em parallel kinematics}, {\em i.e.} with the initial
nucleon momentum $\vec{k}$ parallel to the momentum transfer $\vec{q}$ (most 
available data have been taken in (nearly) perpendicular
 kinematics).  This study has also shown that  multi-step
 processes have a small impact at  large momentum
transfer.  Similar observations could be drawn from a recently published 
(e,e'p) experiment performed at $^4$He \cite{Leeuwe01}. 
This Letter describes the results of the first experiment designed explicitly 
to study  SRC via a measurement of the strength at large $k$ {\em and} $E$ 
under optimal kinematics.

\begin{figure}
\includegraphics[width=7.8cm,clip]{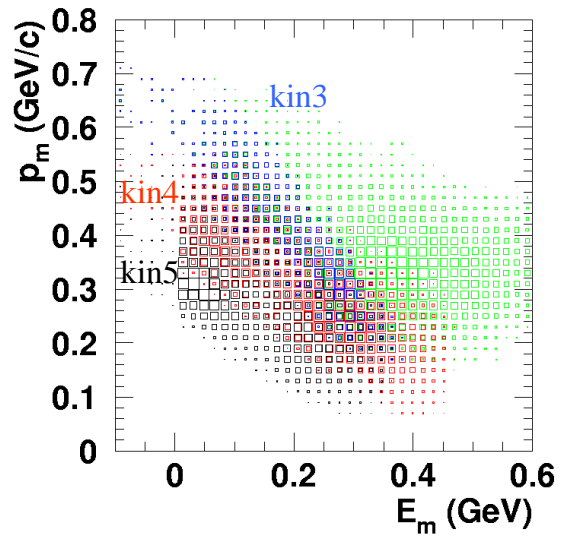}
\caption{\label{empm}Coverage of the $E_m$,$p_m$-plane by the runs taken in
parallel kinematics shown in a cross section times phase space plot.(Due to the large momentum acceptance of the spectrometers,
part of the data (green) are for $\theta_{kq}>45^\circ$).}
\end{figure}

{\bf Experiment.}
The experiment was performed in Hall C at Jefferson Lab employing three
quasi-parallel and two perpendicular kinematics at a q $\gtrsim$ 1~(GeV/c) 
(for a detailed discussion see 
\cite{Rohe04}). Electrons of 3.3~GeV energy 
 and beam currents up to 60~$\mu$A were incident upon  
 $^{12}$C, $^{27}$Al, 
$^{56}$Fe and  $^{197}$Au targets (in the present Letter we limit the discussion 
to $^{12}$C).  The scattered electrons were detected in the HMS spectrometer
(central momenta 2 - 2.8~GeV/c), the protons were detected in the SOS
spectrometer (central momenta 0.8 - 1.7~GeV/c).  Fig.~\ref{empm} gives the
kinematical coverage for the parallel kinematics.  

Data on Hydrogen  were taken as check, 
to determine the various kinematical offsets and to verify the reconstruction 
of particle trajectories and the 
normalizations. Data for the  IP region were also taken.
The resulting proton transparency
agrees with previous determinations \cite{Garrow02} and
modern calculations \cite{Benhar04,Pandharipande92}. The overall accuracy of 
the resulting cross sections is $\pm$6\%.

The spectra of all important observables have been compared to the results of
the Monte Carlo simulation package SIMC of the Hall C collaboration; 
excellent agreement is found. The comparison also shows that the 
resolution  in $E_m$ ($p_m$) is 5~MeV (10~MeV/c).
 
The raw data were analyzed using two different procedures, both based on an
iterative approach and a model spectral function. In one, the phase
space is taken from a Monte Carlo simulation of the experiment, and the spectral
function is determined from the acceptance corrected cross sections. 
Radiative corrections are taken into account according to
\cite{Ent01}. The approach has been verified on special sets of data where 
radiative corrections are large. The other is based on a bin-by-bin comparison of
experimental and Monte Carlo yield, where the Monte Carlo simulates 
the known radiative processes, multiple scattering and energy loss of the
particles, spectrometer transfer matrices, focal plane detector efficiencies,  
the software cuts applied {\em etc}. The parameters of the model spectral
function  then are iterated 
to get agreement between data and simulation.  We have found good agreement 
between the two procedures.

The resulting $S(k,E)$ at low $k, E$ shows the familiar features known from
low-$q$ $(e,e'p)$ experiments
\cite{Lapikas00}. At large $k,E$ we observe the tail resulting from SRC. At 
very large
missing energy $E_m$, the peak due to multi-step interactions involving pion
emission from the various nucleon resonances, appears.  The data taken in
perpendicular kinematics lead to a three times larger strength compared to the
parallel kinematics, which makes it clear that the cross sections measured in perpendicular kinematics receive
dominant contributions from multi-step reactions (the most important ones being
knock-out of another nucleon by the outgoing proton, and processes involving
meson production);
such data then are hardly usable to determine the correlated strength, but can
serve to check our ability to predict multi-step processes.

The $(e,e'p)$ data at low momentum transfer (leading to knock--out protons 
with low momenta $k'$) have
generally been analyzed using a DWBA description for the outgoing proton. At
very large $k'$ the effect of the real part of the optical potential is small,
 particularly for the continuum strength, where a small shift in $k'$ 
is of little concern due to  energy/momentum dependences which are weak as
compared to the ones in the IP region. The main final state interaction effect is the absorption
of the outgoing proton, which is taken into account via the transparency factor
\cite{Benhar04}. For the analysis of the Carbon data, we use
T = 0.60. Also important at large $E$ is the consideration of 
recoil-protons which result from  2-step processes (see below).  

\begin{figure}[t!]
\includegraphics[width=8cm,clip]{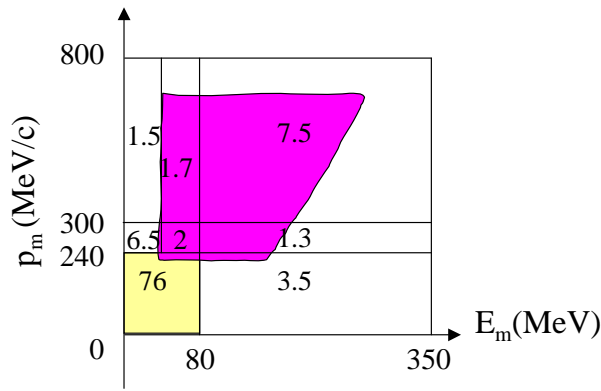}
\caption{\label{integ}Breakdown of the strength from CBF theory in various
integration regions. Numbers in percent. The shaded area is used to determine  the correlated
strength accessible in this experiment. The region labeled '76' contains
the  IP  plus a fraction of the correlated strength.}
\end{figure}
   
{\bf Results.}
Here, we concentrate on the overall strength in the correlated region. 
Fig. \ref{integ} gives, for Correlated Basis Function theory (CBF),  
a schematical breakdown of the various regions of 
interest in the missing energy  $E_m$ and the missing momentum $p_m$ plane, the
quantities that are experimentally defined and identifiable --- in PWIA --- with
$k, E$. The strength corresponding to the IP motion at low $k, E$ 
 amounts to $\sim$80\% for
the CBF calculation \cite{Benhar89}. In some of the regions IP and SRC strength
overlap and cannot be separated. In the shaded region, the strength from SRC is
measurable with the kinematics employed in the present experiment. The shaded
region at large $E_m$ is bounded by a cut that excludes
unwanted contributions from $\Delta$-excitation and $\pi$-production. These
processes have been modeled using MAID \cite{Drechsel99} 
to study possible contributions in our region of interest. 

\begin{figure}[thb]
\includegraphics[height=8cm,angle=270,clip ]{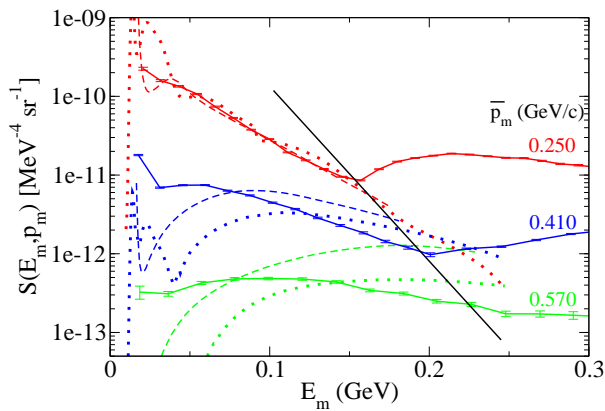}
\caption{\label{comp}Comparison of the experimental spectral function (solid) with the
theories of ref. \cite{Benhar90} (dashed) and ref. \cite{Muther95} (dotted) for three
momentum bins. The line indicates the cut made to separate the correlated and resonance
region. }
\end{figure}

In this shaded region, we find the strength listed in Table \ref{strength}. It
is compared to the strength predicted by theory and integrated over the same
region of $k,E$. This comparison is slightly dependent on the the limits of the 
shaded area as the $k$ and $E$-dependence of
experimental and theoretical $S(k,E)$ are not the same (s. Fig. \ref{comp}); for 
the present comparison we will ignore this minor effect. 

The result shown in Table~\ref{strength} has been obtained using the off--shell
e-p cross section $\sigma_{CC}$ \cite{Rohe04}; for this treatment
 the best agreement of the resulting $S(k,E)$ from different kinematics
(kin3, kin4, kin5) is found. The uncertainty quoted 
includes an estimate for the uncertainty  due to  the
off-shell cross section (judging from difference of strength obtained using 
the cross
sections $\sigma_{CC1}$ and $\sigma_{CC2}$ of \cite{Forest83}). The error does
not contain an uncertainty for the transparency factor used to correct for FSI
 because this value is commonly accepted and in agreement with the 
Glauber calculations of several authors. The statistical error is negligible.

\begin{table}[htb]
\begin{center}
\begin{tabular}{l|l}
\hline
\rule{0mm}{5mm} Experiment & ~~0.61 $\pm 0.06$ \\
~Greens function  theory \cite{Muther95}~~ & ~~0.46 \\
~CBF theory \cite{Benhar89} & ~~0.64 \\[2mm]
\hline 
\end{tabular}
\end{center}
\caption{\label{strength} Correlated strength, integrated over shaded area of 
Fig.\ref{integ} (quoted in
terms of the number of protons in $^{12}$C.)}
\end{table}

For the kinematics of Fig.\ref{empm} the dominant multi-step process is
rescattering of the knocked out nucleon by another nucleon. Barbieri
\cite{Barbieri03a,Barbieri03b} has calculated this process using Glauber theory
and an in-medium N-N cross section accounting for Pauli blocking.
He finds, in agreement with our data, that the multi-step contribution is much
smaller for parallel kinematics. For the experimental result quoted in
Table~\ref{strength},  a multi-step correction of $-$4\% has to be
applied.

{\bf Theoretical predictions of correlated strength.} 
We compare our results to two calculations, performed for $^{16}$O (and adjusted
to correspond to $^{12}$C) and nuclear 
matter, respectively.  For {\em short range} properties, corresponding to 
$S(k,E)$ at large $k$
and $E$,  the results are not expected to depend sensitively on the exact
nucleus \cite{Sick94}.

 
M\"uther {\em et al} \cite{Muther95} 
use the Green's function approach, approximating the
self-energy of the nucleon including all contributions up to second order in
the G-matrix. This finite-nucleus calculation has been performed for  the 
Bonn B  N-N potential.    

In the variational  CBF theory  of
Benhar {\em et al.} \cite{Benhar90} the
correlations are introduced in the wave function via pair-correlation operators
that have the same structure as the terms occurring in the N-N 
interaction.  The Urbana V14+TNI interaction was
used, and the results for $^{12}$C were computed in the local density
approximation \cite{Sick94}.

The theoretical and experimental results are compared in Table \ref{strength}. 
One should note that the strength listed in Table \ref{strength} is only a {\em
fraction} of the total correlated strength; the integral over the correlated
strength covers only the region where it experimentally can be identified 
without undue
contributions from more complicated reactions. The theoretical calculations show
that the correlated strength contributes also at  {\em low} $k$ and $E$, but
there it experimentally cannot be separated from the IP strength.  
The {\em full} correlated
strength amounts to 0.72(1.32) correlated protons  for the Green's function 
 (CBF) approach.  

Table \ref{strength} indicates that 10~\% of the protons could be
found experimentally in the shaded area of Fig. \ref{integ}.
The prediction of CBF theory --- which
overall predicts 22\% of correlated nucleons ---  is close
to experiment, although the detailed $k$ and $E$ dependence does show
deviations; the correlated strength found using the Green's function approach
appears to be somewhat low. This may be related to the fact that the latter approach
employs  a  softer N-N potential which is known to lead to a significantly 
smaller correlation hole in the N-N wave function and to less
correlated strength \cite{Muther00}. A recent finite--nucleus calculation for
$^{12}$C yields a larger correlated strength (16~\%) \cite{Muther04}.

In conclusion, we have performed the first experiment to directly measure the 
strength due to short-range correlations in $(e,e'p)$ reactions at high $(E_m, p_m)$
and have found the results to be in reasonable agreement with predictions. 
  
{\bf Acknowledgments.}
The authors want to thank O. Benhar, C. Barbieri and H. M\"uther for
providing results  and  for many  discussions.
This work was supported by the Schweizerische Nationalfonds, the US Dept. of
Energy  and the US National Science Foundation.

\end{document}